\documentstyle[preprint,revtex]{aps}

\begin{document}
\thispagestyle{empty}


\begin{center}
\hfill{IP-ASTP-08-93}\\
\hfill{December 1993}\\
\hfill{(Revised version)}

\vspace{1 cm}

\begin{title}
Large-scale Polarization of the \\Cosmic Microwave Background Radiation
\end{title}
\vspace{1 cm}

\author{Ka Lok Ng and Kin-Wang Ng}
\vspace{0.5 cm}

\begin{instit}
Institute of Physics, Academia Sinica\\ Taipei, Taiwan 115, R.O.C.
\end{instit}
\end{center}
\vspace{0.5 cm}

\begin{abstract}
The anisotropy and polarization of the cosmic microwave background
radiation (CMBR) induced by the scalar and tensor metric perturbations
are computed in the long-wavelength limit.
It is found that the large-scale polarization of
CMBR induced by the decaying tensor mode can reach a few percents.
This is different from the scalar or inflation-induced tensor mode,
whereas the polarization is at least two orders of magnitude lower.
The effect of matter re-ionization on CMBR is also considered.
We conclude that measuring
the polarization of CMBR on large-angular scales can
probe the ionization history of the early Universe, and test different
cosmological models.

\vspace{0.5 cm}
\end{abstract}
\newpage


The recent detection of large-angular-scale temperature quadrupole anisotropy
in the cosmic microwave background radiation (CMBR) by the DMR aboard the
COBE satellite opens a window to our understanding of physics associated
with the initial conditions of the early Universe.  The most important
conclusion to be drawn from the COBE data on the temperature correlation
function is that they are consistent with a scale-invariant spectrum
of primordial density perturbations \cite {smoot} predicted by
inflationary cosmology \cite {hawking}.

There are two sources of CMBR anisotropy, namely, the density
perturbations (scalar mode) and the primordial gravitational waves \cite{wise}
(tensor mode).  As density perturbations and gravity waves enter the horizon,
they induce large-scale temperature anisotropies in the CMBR via the
Sachs-Wolfe (SW) effect \cite {sachs}.
It is generally believed that the scalar mode can account for the CMBR
anisotropy, which is regarded as evidence for the seeds of galaxies formation.

Recently, it was argued that the anisotropy might
be dominated by the tensor mode \cite {krauss,white}, and was showed that the
tensor-mode dominance actually occurs in certain inflation models \cite
{davis}.
In Ref. \cite {davis}, the authors suggested that by comparing large- and
small-scale anisotropy measurements, one can separate the scalar- and
tensor-mode contributions.  At small-angular scales ($\simeq 1^o$),
the tensor-mode contribution
to the CMBR anisotropy decreases relative to the scalar mode.
A preliminary measurement of the CMBR anisotropy on $1^o$ have indicated
a low value when assuming a scalar-mode contribution \cite {gaier}. If the
result should
hold true, then this may be an evidence for the tensor-mode dominance.
However, an alternative explanation is that late matter re-ionization
may wash out the small-scale anisotropy too.  A potential method for
distinguishing tensor mode from re-ionization scenario is to measure the
polarization component of the CMBR on small-angular scales: re-ionization
leads to much greater polarization.  Anisotropic radiation
acquires linear polarization when it is scattered with free electrons (Thomson
scattering) \cite {rees}.  The effect of Thomson scattering
on the polarization of CMBR for the scalar-mode and tensor-mode cases
can be found in Refs. \cite {bonds} and \cite{pol} respectively.
Ref. [10] showed that roughly 10$\%$ of CMBR anisotropy is polarized
on angular scales less than 1$^o$. Ref. [11] showed that tensor mode
induces large polarization at small angular scales if primordial
spectrum extends to small scales.
In this paper we estimate the
polarization to anisotropy ratio of CMBR due to long-wavelength scalar and
tensor modes in an universe with and without a late re-ionization era.
The result shows that polarization measurements of CMBR on large-angular
scales can probe the reionization history of the early Universe and
possibly test different cosmological models.

To study how polarized photons propagate in the Universe, one need to
solve the equation of transfer for photons.  The general formalism for
the subject of radiative transfer is given in Ref. \cite {chan}.  In general,
arbitrarily polarized photons are characterized by four Stokes parameters,
${\bf n}=(n_l,n_r,n_u,n_v)$, where $n=n_l+n_r$ is the distribution
function for photons with $l$ and $r$ denoting two directions
at right angle to each other.

We shall use the units $c=\hbar=1$ throughout.
The metric that we use is of the flat Robertson-Walker form

\begin{equation}
ds^2=a^2(\eta) \left(d\eta^2-d{\bf x}^2 \right) \;,
\label{e1}
\end{equation}
where $a(\eta)$ and $d\eta=dt/a(t)$ are the scale factor and conformal
time respectively.
Then, it follows that, for a matter-dominated universe,

\begin{equation}
p=0,\;\;a(\eta)={{2\eta^2}\over {H_0}},\;\;\rho={{3H_0^2}\over
{8\pi G \eta^6}},\;\;\eta_0=1,
\label{e2}
\end{equation}
where $p$, $\rho$, and $H_0$ are the pressure, energy density, and
present Hubble constant respectively. Note that we normalize the conformal time
to unity today. In this metric, $\Omega_{\rm total}=\Omega_{DM}+\Omega_B=1$,
where $\Omega_{DM}$ and $\Omega_B$ denote respectively the dark and
baryonic matter.

Then, the equation of transfer for an arbitrarily polarized photon
is governed by the collisional Boltzmann equation,

\begin{equation}
\left(          {\partial \over {\partial \eta}}
      + {\hat e}\cdot{\partial \over {\partial {\bf x}}} \right) {\bf n}
    = - { {d\nu} \over {d\eta} }
         { {\partial {\bf n}} \over {\partial \nu} }
      - \sigma_T N_e a
        \left[ {\bf n} - {1 \over {4 \pi}} \int_{-1}^1 \int_0^{2\pi}
        {P(\mu,\phi,\mu^{'},\phi^{'}) {\bf n} d\mu^{'} d\phi^{'}} \right] \;,
\label{e3}
\end{equation}
where $\sigma_T$ is the Thomson scattering cross section, $N_e$ is
the number of free electrons per unit volume,
($\mu={\rm cos} \theta, \phi$) are the polar angles
of the propagation direction $\hat e$ of the photon with a comoving frequency
$\nu$,
and $P$ is the phase matrix for Thomson scattering given in Ref. \cite {chan}.

The first term on the right-hand side of Eq. (3) describes how
the frequency of a photon is shifted by metric fluctuations.  The second term
accounts for the Thomson scattering effect on photon polarizations.
For an universe with the metric (1),
the solutions of the perturbed Einstein field
equation are given in \cite {sachs}.  In the matter-dominated epoch, the
solution for the spatial part of the metric perturbations is given by

\begin{equation}
h_{ij} = {1 \over \eta} {\partial \over {\partial \eta}}
                         \left( D_{ij} \over \eta \right )
          - 2 \left ({8 \over \eta^3 }- {\nabla ^2 \over \eta } \right)
              \left( C_{i,j} + C_{j,i} \right)
          + {A_{,ij} \over \eta^3} + \eta_{ij}B - {\eta^2 \over 10} B_{,ij}
\label{e4}
\end{equation}
up to a gauge transformation, and the matter density fluctuation

\begin{equation}
\delta \rho = {H^2_0 \over {32\pi G}} \nabla ^2
              \left( {{6A }\over \eta^9} - {{3B \over 5\eta^4}} \right) \;,
\label{e5}
\end{equation}
where the functions $A(\bf {x})$ and $B(\bf {x})$, $C_i(\bf {x})$
and $D_{ij}(\bf{x},\eta)$
correspond to the scalar, vector and tensor metric perturbations respectively.
In the presence of the metric
perturbations, the rate of change of the photon frequency is given by the
SW formula

\begin{equation}
{1 \over \nu} {d\nu \over d\eta}
= {1\over 2} { {\partial h_{ij}}\over {\partial \eta} }e^i e^j
   - {{\partial h_{j0}} \over {\partial \eta}}e^j \;.
\label{e6}
\end{equation}

At a sufficiently early epoch (onset of matter-dominated era at redshifts
$z\le 10^4$, say), photon
radiation is assumed to be a blackbody and unpolarized. Afterwards, the photons
interact with matter through Thomson scatterings, and the subsequent evolution
for photon distribution function is determined by Eq. (3). Solutions for
Eq. (3) which are relevant to the present consideration have been obtained in
two cases: the first case is that the frequency
shift is owing to weak gravitational waves (corresponding to the $D_{ij}$
functions in Eq. (4)) \cite {pol}, and the second due to anisotropic expansion
of an universe \cite {rees,basko}. It was found that in both cases the
polarization to anisotropy ratio of CMBR on large-angular scales is about a few
percents. If matter re-ionization occurs subsequent to the hydrogen
recombination, the ratio will be much enhanced in the anisotropic model
\cite {basko}.
It is realized that the case of unequal Hubble expansion along
three axes is equivalent to the infinite-wavelength limit of the tensor-mode
case up to an overall normalization factor, which is determined by the
magnitude of the gravitational wave \cite {pol}.  Also, it is equivalent to
an axisymmetric universe up to a spherical harmonic function \cite {basko}.
Below we will consider the effect of a generic scalar perturbation to
the large-scale anisotropy and polarization of CMBR,
i.e., setting $D_{ij}=C_i=0$ (hence $h_{j0}=0$) in Eq. (4).
Then, we will consider the large-scale
anisotropy and polarization
of CMBR due to the tensor perturbation (corresponding to $D_{ij}$)
generated in an inflationary cosmology.

We assume a density perturbation wave

\begin{equation}
\delta\rho=\delta_0 \rho \; e^{i{\bf k}\cdot{\bf x}}
\label{e6}
\end{equation}
at the present time $\eta_0=1$ with comoving wave vector ${\bf k}$.
It would induce a small amount of anisotropy and
polarization in the CMBR. Thus we expand the photon distribution function as

\begin{equation}
{\bf n}={\bf n}_0 + {1\over 2} n_0 \; {\bf n}_1 (\eta) \; e^{i{\bf k}\cdot{\bf
x}}\;, \label{e7}
\end{equation}
where $n_0$ is the blackbody distribution function,
${\bf n}_0={1\over 2}n_0 (1,1,0,0)$ denotes unpolarized photons,
and ${\bf n}_1$ is a small fluctuation.
Choosing ${\bf k}$ along with the z-axis, and making use
of the axial symmetry of the problem, we obtain from Eq. (3) the equation of
transfer for the reduced perturbation ${\bf n}_1=({n_1}_l,{n_1}_r)$
(${n_1}_u={n_1}_v=0$) as

\begin{eqnarray}
\left( {\partial \over {\partial \eta}} + ik\mu \right) {\bf n}_1
    = &-& { {\partial {\rm ln} n_0} \over {\partial {\rm ln}\nu} } {1\over \nu}
        { {d\nu} \over {d\eta} } e^{-i{\bf k}\cdot{\bf x}}
        \left( \matrix{1\cr 1\cr} \right)  \nonumber \\
      &-& \sigma_T N_e a
        \left[ {\bf n}_1  - {3 \over 8} \int_{-1}^1
        \left( \matrix{2(1-\mu^2)(1-\mu^{'2})+\mu^2\mu^{'2} & \mu^2 \cr
                        \mu^{'^2} & 1 \cr} \right)
        {\bf n}_1 d\mu^{'} \right] \;, \nonumber \\
\label{e9}
\end{eqnarray}
where $\gamma \equiv {\partial {\rm ln} n_0}/{\partial {\rm ln}\nu}=-1$ in the
Rayleigh-Jeans region.

To evaluate $\nu^{-1}d\nu/d\eta$ in Eq. (9), we separate two cases:
$A\ne 0$
and $B=0$; $B\ne 0$ and $A=0$, corresponding to $\delta\rho/\rho$ which
decreases and increases with time respectively. The increasing mode is
believed to be responsible for the structure formation of the Universe. From
Eqs. (4)-(7), we find that

\begin{equation}
{1\over \nu}{{d\nu}\over{d\eta}}=\cases{-3{{\delta_0\rho}\over
    \rho_0}{1\over \eta^4} \mu^2 e^{i{\bf k}\cdot{\bf x}} & for $A\ne 0\;$;\cr
    2{{\delta_0\rho}\over
    \rho_0} \eta \mu^2 e^{i{\bf k}\cdot{\bf x}} & for $B\ne 0\;$,\cr}
\label{e10}
\end{equation}
where $\rho_0=3H_0^2 /(8\pi G)$ is the present average energy density.
In either case,
Eq. (9) is too complicated to be solved. However, by taking infinite-wavelength
limit, i.e., $k\rightarrow 0$, one could extract simple solutions from Eq. (9).
These solutions are indeed good estimates of the anisotropy and
polarization of CMBR on large-angular scales (see below).

By substituting Eq. (10) in Eq. (9), taking $k\rightarrow 0$ limit (i.e.,
neglecting the term $ik\mu {\bf n}_1$), and performing mode-decomposition,

\begin{equation}
{\bf n}_1= \alpha (\mu^2-{1\over 3})\left( \matrix{1\cr 1\cr}\right)
          +\beta  (1-\mu^2)         \left( \matrix{1\cr -1\cr} \right) \;,
\label{e11}
\end{equation}
where $\alpha$ and $\beta$ are respectively the anisotropy and polarization of
the cosmic radiation at time $\eta$ (in essence, $\alpha$ represents the
temperature anisotropy, $\delta T/T$, and $\beta$ represents the fractional
temperature difference between two orthogonal directions along the line of
sight), we obtain

\begin{eqnarray}
{{d\alpha}\over{d\eta}}&=& -\gamma\Delta(\eta) - \sigma_T N_e a
         \left( {9\over 10}\alpha + {3\over 5}\beta \right)\;, \label{e12}\\
         {{d\beta}\over{d\eta}} &=& - \sigma_T N_e a
         \left( {1\over 10}\alpha + {2\over 5}\beta \right)\;, \label{e13}
\end{eqnarray}
where, in terms of redshift $z$,

\begin{equation}
{{\Delta(\eta)}\over {2\eta^2}} \equiv \Delta(z)
        =\cases{-{3\over 2}{{\delta_0\rho}\over\rho_0}(1+z)^3&$A\ne 0\;$;\cr
                {{\delta_0\rho}\over\rho_0}(1+z)^{1\over 2}&$B\ne 0\;$.\cr}
\label{e14}
\end{equation}
Integrating $\alpha$ and $\beta$ in Eqs. (12) and (13) with respect to $z$,
we obtain their present values

\begin{eqnarray}
\alpha&=&-\gamma \int_0^{\infty} \Delta(z) \left[{6\over 7}
                 e^{-\tau(z)}+{1\over 7}e^{-{3\over 10}\tau(z)}\right]
                 (1+z)^{-{5\over 2}}dz \label{e15} \\
          \beta&=&-\gamma \int_0^{\infty} \Delta(z) \left[{1\over 7}
                 e^{-\tau(z)}-{1\over 7}e^{-{3\over 10}\tau(z)}\right]
                 (1+z)^{-{5\over 2}}dz \label{e16}
\end{eqnarray}
where $\Delta(z)$ is given in Eq. (14) and the optical depth is given by

\begin{equation}
\tau(z)=\tau_0 \int_0^z \chi_e(z') \sqrt{1+z'} dz'\; ;\;\;\;
\chi_e \equiv {N_e \over N_B}\;,\;\;\tau_0={{\sigma_T \rho_c \Omega_B}\over
{H_0 m_B}}\;,
\label{e17}
\end{equation}
where $\chi_e$ is the degree of ionization, $m_B$ and $N_B$ are respectively
the proton mass and number density, $\rho_c$ is the critical energy density,
and $\Omega_B \equiv \rho_B/\rho_c$. Henceforth, we assume $H_0=100h\;
{\rm km \;sec^{-1} Mpc^{-1}}$ with $h=1$. Then, $\tau_0=0.0696\;\Omega_B h$.

To evaluate $\alpha$ and $\beta$ in Eqs. (15) and (16), we need to determine
the optical depth which is determined by the dynamics of hydroden recombination
and the
history of re-ionization. Let us first assume an instantaneous recombination:
$\chi_e=1$ for $z\ge z_r$ and $\chi_e=0$ for $z< z_r$, where $z_r$ is the
redshift at which hydrogen recombination occurs; $z_r=1350, 1450, 1500$ for
$\Omega_B=0.01, 0.1, 1$ respectively.
Express $\alpha=\pm \alpha_0 \gamma \; \delta_0 \rho/\rho_0$,
where the plus (minus) sign corresponds to the $A\ne 0$ ($B\ne 0$) case.

For $A\ne 0$ we obtain from Eqs. (14)-(17), $\alpha_0 \simeq 5 \times 10^4$
and ${\beta / \alpha} \simeq 10^{-2}-10^{-4}$ for $\Omega_B$
from 0.01 to 1. In this case, $\alpha
\simeq -0.5$ when assuming a horizon-sized density perturbation with amplitude
$\delta_0 \rho/\rho_0 \simeq 10^{-5}$, which is too big when compared to
observational data.
However, the $A \ne 0$ case is equivalent to
(i) an axisymmetric universe up to a numerical factor, and (ii)
that of an infinitely long decaying gravitational wave up to a
numerical factor times a
spherical harmonic function. Nevertheless, all the three cases have the same
${\beta / \alpha}$ ratio.
The reasons for the above equivalences are the following.
For an axisymmetric universe with the present shear given by $\Delta H_0$, the
rate of change of the photon frequency is given by \cite{basko}

\begin{equation}
{1\over \nu}{{d\nu}\over{d\eta}}={{2\Delta H_0}\over{H_0}}{1\over
\eta^4}(\mu^2-{1\over 3})\;,
\label{e18}
\end{equation}
which is similar to the $A \ne 0$ case in Eq. (10) ($k\rightarrow 0$)
up to an irrelevant monopole term. In the presence of a monochromatic
gravitational wave with comoving wave vector ${\bf k}$ and polarization
$\lambda=+$ mode, it was shown that \cite{pol}

\begin{equation}
{1\over \nu}{{d\nu}\over{d\eta}}={1\over 2}(1-\mu^2)\;{\rm cos}2\phi\;e^{-i{\bf
k}\cdot{\bf x}}{d\over {d\eta}}(h\;e^{ik\eta})\;;\;\;
h={ikf \over{\eta^2}} \left ( 1+{i\over{k\eta}} \right)\;,
\label{e19}
\end{equation}
where $f$ is a constant wave amplitude. Taking the $k\rightarrow 0$ limit, Eq.
(19) is similar to the $A \ne 0$ case in Eq. (10).  For such wave, the mode
decomposition is given by

\begin{equation}
{\bf n_1}=  {\alpha \over 2} ({1-\mu^2} ){\rm cos}2\phi
                             \left( \matrix{1\cr 1\cr 0\cr}\right) +
            {\beta  \over 2} \left( \matrix{(1+\mu^2){\rm cos}2\phi\cr
              -(1+\mu^2){\rm cos}2\phi\cr 4\mu {\rm sin}2\phi\cr} \right) \;.
\label{e20}
\end{equation}

For $B\ne 0$,
we find $\alpha_0 \simeq 1$ and ${\beta / \alpha} \simeq 10^{-5}-10^{-7}$ for
$\Omega_B$
from 0.01 to 1.
Then, $\alpha \simeq \delta_0\rho/\rho_0 \simeq 10^{-5}$ which is the
well-known
SW contribution to the large-scale anisotropy of CMBR. $\beta$ is below
$10^{-10}$ which is consistent with the values of lower multipole moments that
were given in Ref. \cite{bonds}.

To determine the effects of non-instantaneous recombination, we approximate the
history of ionization by,

\begin{equation}
\chi_e=\cases{3.2\times 10^{-4} \Omega_B^{-1}&$z<900\;$,\cr
              5.68\times 10^6 \Omega_B^{-1} z^{-1} e^{-{14453\over z}}&
              $900<z<1500\;$,\cr 1&$z>1500\;$,\cr}
\label{e21}
\end{equation}
where we assume that a residual ionization remains for $z<900$.  For
$900<z<1500$,
we use the results given in Ref. \cite {sun} which take into account the
possibility of two-photon decay of the hydrogen 2S level during the process of
recombination. For $z>1500$, we assume a complete ionization (an exact
treatment
would require using the Saha equation). Plugging Eq. (21) in Eq. (17) and then
using Eqs. (14)-(16), we find that the degree of anisotropy, on comparing with
the instantaneous case, remains approximately the same, whereas the degree of
polarization can possibly greatly enhanced. Being weakly dependent on the value
of $\Omega_B$, the ratio ${\beta / \alpha}$ increases to about $-4\%$ for
the case of $A\ne 0$
and $-10^{-4}$ for that of $B \ne 0$. These results are shown in Figs. 1 and 2.
Our value of the degree of polarization for the $A\ne 0$ case is consistent
with the $2\%$ result that was given in Ref. \cite{pol}. To understand our
results for the non-instantaneous case, we examine the value of the optical
depth evaluated at the beginning
of hydrogen recombination. This value denotes the number of Thomson
scatterings between epochs $z=z_r$ and $z=0$. We find that $\tau(z=1500)\simeq
90$, which
is much larger than one. This explains why the polarization is sensitive to the
recombination dynamics.

Finally, we consider the effects of matter re-ionization subsequent to the
recombination epoch. This re-ionization of matter may attribute to the
reheating by radiation
released from protogalaxies during galaxy formation. To demonstrate this, we
approximate the ionization fraction by $\chi_e=1$ for $z>z_r$; 0 for
$z_h<z<z_r$; 1 for $z<z_h$, where $z_h$ is the redshift at which the
re-ionization occurs. Note that the case of instantaneous recombination is
equivalent to setting $z_h=0$.  Using Eqs. (14)-(17) and varying $z_h$ from
10 to 100, we find that
the polarization, being very sensitive to the re-ionization, becomes much
larger (see Fig. 2). To understand why the polarization curves all suddenly
change around $z_h=10$, we evaluate the optical depth at $z=10$. We find that
$\tau(10)\simeq 1.6 \Omega_B$. This explains the jumps for all the cases
except the $B \ne 0$ and $\Omega_B=0.01$ case. However, in this case
the polarization before recombination is extremely small, even a small
amount of scatterings during the re-ionization epoch can induce an increase in
the polarization over its prerecombination value. As for the anisotopy
$\alpha_0$, it is found that it remains almost unchanged in the case of $B \ne
0$ and it is diminished in that of $A \ne 0$ (see Fig. 1). As mentioned above,
the $A \ne 0$ case is equivalent to an axisymmetric universe, this decreasing
behavior agrees with the results given in Ref. \cite{basko}.

We are now to justify the infinite-wavelength approximation in our
calculations.
Since horizon-sized waves dominate
the contributions to the anisotropy and hence polarization of CMBR
on large-angular scales via the SW effect either for the scalar \cite
{pee} or tensor mode \cite {white}.  Thus, one should compare the
Thomson scattering mean-free path,
$(\sigma_T N_e a)^{-1}$, with the wavelength of horizon-sized density
perturbations, $k^{-1}_{hor}$, at the moment when the contributions to the
integrals in Eqs. (15) and (16) are dominant.
If the former is much smaller than the latter, then the term $ik\mu{\bf n}_1$
in Eq. (9)
can be neglected. For example, in the case of instantaneous recombination, the
polarization
$\beta$ receives contribution mainly from $z \ge z_r$. For $\Omega_B=0.01$ and
$z_r=1350$, $\sigma_T N_e a \ge 2500$ for $z \ge z_r$ which is much
bigger than $k_{hor}\simeq 1$. For other values of $\Omega_B$ and $z_h$,
we find that $\sigma_T N_e
a$ is at least comparable to or much bigger than 1. Thus, we conclude
that taking the infinite-wavelength limit is an efficient approximation
in calculating the lower multipole moments, especially the quadrupole
moment. However, the finite-wavelength correction is of
interest for intermediate-scale calculation.

We now turn to the inflation-induced tensor mode calculation.
The calculation is similar to the case of decaying tensor mode. The rate of
change of the photon frequency is given by Eq.(19) with $h e^{ik\eta}$
replaced by $3 A(k)j_1(k\eta) /(k\eta)$, where
$A^2(k)=8v/(3\pi)$ is the scale invariant spectrum, $j_1(k\eta)$ is the
spherical Bessel function of order 1, and $v$ is a parameter depending on the
inflation scale \cite {wise,krauss,white}.  By taking the infinite-wavelength
limit, the rate of change of the photon frequency decreases to zero. This is
expected since the amplitude of the tensor mode well outside the horizon is
constant. Therefore, the above limiting approach fails in this case. However,
as we have mentioned that horizon-sized waves dominate the contributions,
we instead proceed the calculation by simply dropping the term $ik\mu{\bf n}_1$
and taking $k=2\pi$ in Eq. (9), and using $v=4 \times 10^{-11}$.
The results are shown in Figs. 1 and 2.
The behavior of the results is very similar to that of the $B \ne 0$ case.
This makes it difficult to distinguish the scalar mode from the
inflation-induced   tensor mode by measuring the large-scale anisotropy and
polarization of CMBR.
We have also carried a full numerical calculation of Eq. (9) by taking
$k\simeq 1$ and found that dropping the term $ik\mu{\bf n}_1$ does not make
any significant difference.

In conclusion, we have calculated the large-scale anisotropy and polarization
of CMBR due to the scalar and tensor metric perturbations,
as well as the effects
of the non-instantaneous recombination and matter re-ionization.
The results are shown in Figs.1 and 2.  The $\alpha_{0}$ (up to an
overall normalization factor which is determined by the amplitude of the
decaying tensor mode) and ${\beta / \alpha}$ values for the
$A\ne 0$ case also represent respectively the anisotropy and the polarization
to anisotropy ratio due to the decaying tensor mode.  Although it seems
unlikely that measurable super-horizon sized decaying tensor modes could be
generated in the early Universe, however, the large polarization due to this
decaying-type perturbations or anisotropic Hubble expansion \cite{zeldovich}
remains an interesting possibility.
It is shown that measuring the degree
of polarization of CMBR on large-angular scales is a potential method for
probing the ionization history of
the early Universe. The linear polarization of CMBR was measured at 33 GHz
over 11 declinations from $37^o$S to $63^o$N. Fitting the data to
an axisymmetric model and spherical harmonics through the third order yields
$|\beta|<3\times 10^{-5}$ and $|\beta|<6\times 10^{-5}$ respectively at $95\%$
confidence level \cite {lubin}. These upper limits can hardly be used to put
any
constraint on the present calculations. Currently, the linear antenna aboard
COBE has collected data for the polarization component of CMBR which is being
analysed.  Any trace of polarization or a better limit will prove invaluable
to our understanding of the early Universe.

\vspace{3 cm}
\noindent{{\bf Acknowledgements}}
\vspace{0.5 cm}

This work was supported in part by the R.O.C. NSC Grant No.
NSC82-0208-M-001-059, NSC83-0208-M-001-053 and NSC82-0208-M-001-131-T.

\newpage

\centerline{{\bf FIGURE CAPTIONS}}
\vspace{1 cm}

\noindent{Figure 1. Large-scale anisotropy of CMBR versus the redshift at which
matter re-ionization occurs.
Symbols $A$, $B$ and $T$ denote respectively the cases of $A\ne 0$, $B\ne 0$
and
the inflation-induced tensor. For the cases of $A\ne 0$ and $B\ne 0$,
$\alpha=\pm \alpha_0 \gamma \;\delta_0\rho/\rho_0$. For the tensor case,
$\alpha=\gamma\alpha_0 10^{-6}$.
$z_h=0$ corresponds to instantaneous recombination.
Each solid triangle corresponds to non-instantaneous recombination for
$\Omega_{B} =1, 0.1 \; {\rm and} \;0.01$. Dashed, solid
and dotted curves correspond respectively to
$\Omega_B=1, 0.1 \;{\rm and} \;0.01$.}

\vspace{0.5 cm}
\noindent{Figure 2. Large-scale polarization to anisotropy ratio of CMBR.
Symbols $A$, $B$ and $T$ denote respectively the cases of $A\ne 0$, $B\ne 0$
and
the inflation-induced tensor.
$z_h=0$ corresponds to instantaneous recombination.
Each solid triangle corresponds to non-instantaneous recombination for
$\Omega_{B} =1, 0.1 \;{\rm and} \;0.01$. Dashed, solid
and dotted curves correspond respectively to
$\Omega_B=1, 0.1 \;{\rm and} \;0.01$.}

\end{document}